\documentclass[aps,prl,10pt,twocolumn,groupedaddress,showpacs,floatfix]{revtex4-1}
\usepackage{dcolumn}
\usepackage{amsmath}
\usepackage{bm}
\usepackage{times}
\usepackage{graphicx}
\usepackage{xcolor}
\newcolumntype{.}{D{x}{}{-1}}
\newcommand*{\cent}[1]{\multicolumn{1}{c}{$#1$}}
\newcommand*{\centt}[1]{\multicolumn{1}{c}{#1}}
\newcolumntype{w}[1]{D{.}{.}{#1}}

\begin{document}
\preprint{Version 1.1}

\title{Complete $\alpha^6\,m$ corrections to the ground state of H$_2$}
\author{Mariusz Puchalski} 
%\email{mpuchals@amu.edu.pl}
\affiliation{Faculty of Chemistry, Adam Mickiewicz University,
             Umultowska 89b, 61-614 Pozna{\'n}, Poland}
\author{Jacek Komasa} 
%\email{komasa@amu.edu.pl}
\affiliation{Faculty of Chemistry, Adam Mickiewicz University,
             Umultowska 89b, 61-614 Pozna{\'n}, Poland}
\author{Pawe\l\ Czachorowski}
%\email{pczachor@fuw.edu.pl}
\affiliation{Faculty of Physics, University of Warsaw, Pasteura 5, 02-093 Warsaw, Poland}
\author{Krzysztof Pachucki}
%\email{krp@fuw.edu.pl} 
\affiliation{Faculty of Physics, University of Warsaw, Pasteura 5, 02-093 Warsaw, Poland}

\date{\today}

\begin{abstract}
We  perform the calculation of all relativistic and quantum electrodynamic
corrections of the order of $\alpha^6\,m$ to the ground electronic state of 
a hydrogen molecule and present improved results
for the dissociation and the fundamental transition energies.
These results open the window for the high-precision spectroscopy of H$_2$
and related low-energy tests of fundamental interactions.
\end{abstract}

\maketitle

The hydrogen atom and various hydrogenic systems like positronium, muonium, 
muonic hydrogen, and He$^+$, due to highly accurate theoretical predictions \cite{eides},
are considered for the determination of fundamental 
physical constants \cite{nist} and for the low-energy tests of the Standard Model \cite{crivelli, pohl}.
However, they are limited by uncertainties in the nuclear structure
or natural life-time of the system.
The $1S-2S$ transition in H is the best example,
where the precision of the measurement $f(1S-2S) = 2\,466\,061\,413\,187\,035(10)$ Hz
\cite{hydrogen} exceeds by orders of magnitude
any theoretical predictions. This is because of the relatively large theoretical 
uncertainties in the proton structure and resulting inaccuracies in fundamental constants.
The lack of another sharp transition in the hydrogen 
makes the determination of the Rydberg (R$_\infty$) constant, 
which transforms atomic units to inverse of the transition wavelength, 
much less accurate than it would be if another such transition was available. 
Here we point out that the dissociation energy of H$_2$ can serve this purpose, 
as it is stable in the ground electronic state and can be calculated 
with sufficient precision. So having two accurate and calculable transitions
the two unknowns R$_\infty$ and $r_p$ can be determined,
which among others, would help resolve the proton charge radius puzzle. 
Another alternative systems for which high precision calculations 
are possible, include the helium ion He$^+$ \cite{eikema}, heavy hydrogen like 
ions \cite{ulj}, and the hydrogen molecular ion \cite{korobov, h2plus}.
 
The calculations for the hydrogen molecule have never been considered 
to be as accurate as for the hydrogen atom due to the lack of an analytic solution
of the Schr\"odinger equation. However, the numerical solution of this equation,
as has been shown recently \cite{nonad}, can be as accurate as $10^{-12}$,
and thus it will not limit the accuracy of theoretical predictions.
There are obviously various corrections, such as relativistic and quantum electrodynamic
(QED) ones. So far, they have been calculated up to $\alpha^5\,m$ order \cite{piszcz},
and only in the adiabatic approximation. Beyond this approximation, namely the combined 
nonadiabatic and relativistic effects,
have not yet been obtained and they will limit the accuracy of current predictions.
Here we calculate one of the most difficult, the $\alpha^6\,m$ correction,
using the so-called nonrelativistic QED approach. Next, we point out
that when the higher order $\alpha^7\,m$ correction is determined,
energies of the hydrogen molecule can be obtained almost as accurately as those of the hydrogen atom alone,
and thus may be used for determination of the R$_\infty$ constant.
Meanwhile, on the basis of the $\alpha^6\,m$ correction obtained herein, 
we will present improved results 
for the dissociation and the fundamental transition energies.
 
\paragraph{NRQED effective Hamiltonian}
Since there is no formulation of QED theory based on a multielectron Dirac equation
with Coulomb interactions, we use an effective nonrelativistic QED (NRQED) approach that is based on 
the Schr\"odinger equation. 
According to QED theory, the expansion of energy levels in powers of 
the fine structure constant $\alpha$ has the following form 
\begin{equation}
E(\alpha) = E^{(2)} + E^{(4)} + E^{(5)} + E^{(6)} + E^{(7)} + O(\alpha^8), \label{01}
\end{equation}
where $E^{(n)}$ is a contribution of order $\alpha^n\,m$ and may include 
powers of $\ln\alpha$. 
Each expansion term $E^{(n)}$ can be expressed as an expectation
value of some effective Hamiltonian with the nonrelativistic wave function $\Phi$. 
The first one, $E^{(2)}\equiv E_0$, is the eigenvalue of the
nonrelativistic Hamiltonian $H_0$. 
In the infinite nuclear mass approximation (in theoretical units $\hbar = c = 1$)
\begin{equation}
H_0 = \frac{\vec p_1^{\,2}}{2\,m} + \frac{\vec p_2^{\,2}}{2\,m} + V, \label{02}
\end{equation}
where
\begin{equation}
V =        -\frac{Z_A\,{\alpha}}{r_{1A}} - \frac{Z_A\,\alpha}{r_{2A}}
             -\frac{Z_B\,\alpha}{r_{1B}} - \frac{Z_B\,\alpha}{r_{2B}}
                +\frac{\alpha}{r} + \frac{Z_A\,Z_B\,\alpha}{r_{AB}}
\label{03}
\end{equation}
$r = r_{12}$, and where indices 1 and 2 correspond to electrons, whereas $A$ and $B$ correspond to nuclei.
The next term of this expansion $E^{(4)}$ is the expectation value of 
the well-known Breit-Pauli (BP) Hamiltonian $H^{(4)}$ \cite{bs}.
$E^{(5)}$ is the leading QED contribution, which is well defined and 
can also be expressed in terms of matrix elements of somewhat
more complicated operators \cite{sap,piszcz}.
The calculation of the next term in $\alpha$-expansion $E^{(6)}$ is the subject of the present work. 
This term can be represented as
\begin{equation}
E^{(6)} = \bigl\langle H^{(6)} \bigr\rangle + 
\biggl\langle H^{(4)}\,\frac{1}{(E_0-H_0)'}\,H^{(4)}\biggr\rangle, \label{05}
\end{equation}
where $H^{(6)}$ is the effective Hamiltonian of order $\alpha^6\,m$.
Its derivation is presented in the following paragraph. Here,
the second-order contribution, and correspondingly $H^{(4)}$, 
is split into two parts depending on the symmetry of intermediate states.
\begin{equation}
E_A = \biggl\langle H_A\,\frac{1}{(E_0-H_0)'}\,H_A\biggr\rangle\,,\,
E_C = \biggl\langle H_C\,\frac{1}{E_0-H_0}\,H_C\biggr\rangle \label{06}
\end{equation}
where
\begin{eqnarray}
H_A &=& -\frac{p_1^4}{8\,m^3} - \frac{p_2^4}{8\,m^3} 
-\frac{\alpha}{2\,m^2}\,p_1^i\,
\biggl(\frac{\delta^{ij}}{r}+\frac{r^i\,r^j}{r^3}\biggr)\,p_2^j
\nonumber \\ &&
+\frac{\pi\,\alpha}{m^2}\,\delta^3(r)
+\frac{\pi\,Z_A\alpha}{2\,m^2}\,\delta^3(r_{1A})
+\frac{\pi\,Z_A\alpha}{2\,m^2}\,\delta^3(r_{2A})
\nonumber \\ &&
+\frac{\pi\,Z_B\alpha}{2\,m^2}\,\delta^3(r_{1B})
+\frac{\pi\,Z_B\alpha}{2\,m^2}\,\delta^3(r_{2B}),
\label{08}
\end{eqnarray}
and
\begin{eqnarray}
H_C &=& \frac{(\vec\sigma_1-\vec\sigma_2)}{2}\,\biggl[
\frac{Z_A}{4\,m^2}\biggl(\frac{\vec r_{1A}}{r_{1A}^3}\times\vec p_1
-\frac{\vec r_{2A}}{r_{2A}^3}\times\vec p_2\biggr)
\label{09} \\ && \hspace*{-8ex}
+\frac{Z_B}{4\,m^2}\biggl(\frac{\vec r_{1B}}{r_{1B}^3}\times\vec p_1
-\frac{\vec r_{2B}}{r_{2B}^3}\times\vec p_2\biggr)
+\frac{1}{4\,m^2}\,\frac{\vec r}{r^3}\times(\vec p_1+\vec p_2)\biggr].
\nonumber
\end{eqnarray}
% The remaining parts of $H^{(4)}$  do not contribute to the energy of $S=0$ states,
% so they can be omitted for the ground electronic state.
The first term $E_A$ as well as  $\langle H^{(6)}\rangle$ are 
separately divergent, but their sum is finite. We follow the approach of Ref. \cite{hsinglet} 
and use the technique of dimensional regularization to eliminate these divergences
from the matrix elements. $H_A$ in the above equation should therefore be written
in $d$-dimensions, but for simplicity we write only the $d=3$ form.

The  effective Hamiltonian $H^{(6)}$ is derived in an analogous way
as for the He atom in Ref. \cite{hsinglet}. There is no additional complication
for the case of H$_2$, except obviously for the presence of two Coulomb fields
instead of one. It is  expressed as a sum of various contributions, 
%\begin{equation}
$H^{(6)} = H_Q + H_H + H_{R1} + H_{R2}. \label{10}$
%\end{equation}
$H_Q$ is a sum of all terms that come from one- and two-photon exchange
of the low-energy photons $k\sim \alpha\,m$. We do not write their explicit expression because
it is too long. They are divergent at high photon momenta,
or equivalently at small distances $r$ and $r_{aX}$.
This divergence cancel out with the second-order contribution $E_A$
and with the hard three-photon exchange, which in $d=3-2\,\epsilon$ dimensions is \cite{pos}
\begin{equation}
H_H = \biggl(4\ln m -\frac{1}{\epsilon} -\frac{39\,\zeta(3)}{\pi^2}
+\frac{32}{\pi^2}-6\,\ln(2)+\frac{7}{3}\biggr)
\frac{\pi\,\alpha^3}{4\,m^2}\delta^d(r). \label{11}
\end{equation}
Later in  Eq. (\ref{23}) and Table \ref{TBL1} we present a simplified and regularized form of $H_Q$.
The remaining contributions are radiative corrections,
which at the order $\alpha^6\,m$ are proportional to Dirac $\delta$-functions,
and they are known from the hydrogenic case.
The one-loop correction is \cite{eides}
\begin{eqnarray}
H_{R1} &=& \frac{\alpha^3\,\pi}{m^2}\,
\biggl[\frac{427}{96}-2\,\ln(2)\biggr]\,
\bigl[Z_A^2\,\delta^3(r_{1A})+Z_A^2\,\delta^3(r_{2A}) 
\nonumber \\ && \hspace*{5ex}
+ Z_B^2\,\delta^3(r_{1B})+Z_B^2\,\delta^3(r_{2B})\bigr]
\label{12} \\ &&
           +\frac{\alpha^3}{m^2}\,\biggl[
\frac{6\,\zeta(3)}{\pi^2}-\frac{697}{27\,\pi^2}-8\,\ln(2)+\frac{1099}{72}
\biggr]\,\pi\,\delta^3(r), \nonumber
\end{eqnarray}
and the two-loop correction is \cite{eides}
\begin{eqnarray}
H_{R2} &=& \frac{\alpha^3\,\pi}{m^2}\,
\biggl[-\frac{9\,\zeta(3)}{4\,\pi^2}-\frac{2179}{648\,\pi^2}+\frac{3\,\ln(2)}{2}-\frac{10}{27}\biggr]
\nonumber \\ && \hspace*{-7ex}
\times\bigl[Z_A\,\delta^3(r_{1A})+Z_A\,\delta^3(r_{2A}) 
+ Z_B\,\delta^3(r_{1B})+Z_B\,\delta^3(r_{2B})\bigr]
\nonumber \\ && \hspace*{-7ex}
           +\frac{\alpha^3}{m^2}\,\biggl[
\frac{15\,\zeta(3)}{2\,\pi^2}+\frac{631}{54\,\pi^2}-5\,\ln(2)+\frac{29}{27}
\biggr]\,\pi\,\delta^3(r). \label{13}
\end{eqnarray}
At this point we have considered all contributions of the order of $\alpha^6\,m$. 
The higher order term is estimated on the basis of the dominant double logarithmic 
contribution, which for  $Z_A=Z_B=1$ is
\begin{equation}
H^{(7)} \!\approx\! -\frac{\alpha^4}{m^2}\ln^2\bigl(\alpha^{-2}\bigr)
\bigl[\delta^3(r_{1A})+\delta^3(r_{2A}) 
+ \delta^3(r_{1B})+\delta^3(r_{2B})\bigr] \label{14}
\end{equation}

\paragraph{Elimination of Singularities}
The second-order matrix element $E_A$ in Eq. (\ref{06})
requires subtractions of $1/\epsilon$ singularities. 
For this we use the transformation
\begin{equation}
H_A = H'_A +  \bigl\{H_0-E_0,Q\bigr\}, \label{15}
\end{equation}
where
\begin{equation}
Q = -\frac{1}{4}\biggl(\frac{Z_A}{r_{1A}}+\frac{Z_B}{r_{1B}}+\frac{Z_A}{r_{2A}} + \frac{Z_B}{r_{2B}}\biggr)
+ \frac{1}{2\,r}, \label{16}
\end{equation}
so that $E_A = E'_A  + E''_A$, where
\begin{eqnarray}
E'_A &=& \biggl\langle H'_A\,\frac{1}{(E_0-H_0)'}\,H'_A\biggr\rangle, \label{18}\\
E''_A &=& \bigl\langle Q\,(E_0-H_0)\,Q\bigr\rangle
+ 2\,\langle H_A\rangle\,\langle Q\rangle
-\bigl\langle \bigl\{H_A\,,\,Q\bigr\}\bigr\rangle\,.
\nonumber \\ \label{19}
\end{eqnarray}
$E'_A$ is finite in the limit $\epsilon \rightarrow 0$, and
\begin{eqnarray}
H'_A |\Phi\rangle &=& \biggl\{
-\frac{1}{2}\,(E_0-V)^2
-p_1^i\,\frac{1}{2\,r}\,\biggl(\delta^{ij}+\frac{r^i\,r^j}{r^2}\biggr)\,p_2^j
\nonumber \\ &&
+\frac{1}{4}\,\vec \nabla_1^2 \, \vec \nabla_2^2
+ \frac{1}{2}\,(E-V)\,(V_1+V_2) 
\label{20}\\ &&
+ \frac{1}{4}\,\vec\nabla_1\,(V_1+V_2)\,\vec\nabla_1
+ \frac{1}{4}\,\vec\nabla_2\,(V_1+V_2)\,\vec\nabla_2
\biggr\}|\Phi\rangle, \nonumber
\end{eqnarray}
where the action of $\vec \nabla_1^2 \, \vec \nabla_2^2$ on $\Phi$ in the above 
is understood as a differentiation with omission of $\delta^3(r)$, 
and $V_i$ is defined in the caption of Table \ref{TBL1}.

The expression for $E^{(6)}$, after subtraction and elimination of all singularities,
is the main result of this work and has the following form
\begin{equation}
E^{(6)} = E'_Q + E'_H + E'_A + E_C + E_{R1} + E_{R2} -\ln(\alpha)\,\pi\,\langle\delta^3(r)\rangle 
\label{21}
\end{equation}
where $E'_H$ is the expectation value of $H_H$ with dropping $1/\epsilon$ and $\ln m$ terms,
$E'_A$ is defined in Eq. (\ref{18}), $E_C$ in Eq. (\ref{06}), $E_{R1}$, and $E_{R2}$ are
mean values of the Hamiltonians
(\ref{12}) and (\ref{13}), correspondingly.
The logarithmic term in Eq. (\ref{21}) agrees with that obtained for helium in Ref. \cite{lna}.
The sum of the ``soft'' photon exchange contributions $E'_Q = \langle H_Q\rangle + E''_A$
for the case of H$_2$ after $(1 \leftrightarrow 2)$ simplification becomes
\begin{widetext}
\begin{eqnarray}
E'_Q &=& -\frac{E_R^3}{2} + R\,\frac{d E_R}{d R}\,\biggl(\frac{E^{(4)}}{2} - \frac{E_R^2}{4}\biggr) 
- \frac{E_R}{4}\,Q_1 + \frac{1}{8}\,Q_2 - \frac{1}{4}\,Q_3 
- \frac{1}{2}\,Q_4 + \frac{3}{8}\,Q_5 - \frac{1}{4}\,Q_6 
+ \frac{1}{24}\, Q_7 + \frac{2\,E^{(4)} + E_R^2}{4}\,Q_8 
\nonumber \\ &&
- \frac{E_R}{2}\,Q_9 + \frac{1}{4}\,Q_{10} + \frac{E_R}{2}\,Q_{11} 
+ E_R\,Q_{12} - E_R\,Q_{13}
 - Q_{14} + Q_{15} - \frac{1}{2}\,Q_{16} - \frac{1}{2}\,Q_{17} + \frac{1}{16}\,Q_{18} 
+ \frac{1}{2}\,Q_{19} - \frac{1}{8}\,Q_{20} 
\nonumber \\ &&
+ \frac{1}{4}\,Q_{21} + \frac{1}{4}\,Q_{22} + Q_{23} + \frac{1}{2}\,Q_{24} 
- \frac{1}{32}\,Q_{25} - \frac{1}{4}\,Q_{26} - \frac{E_R}{8}\,Q_{27} 
- \frac{1}{2}\,Q_{28} + \frac{1}{4}\,Q_{29} + \frac{1}{8}\,Q_{30}
\label{23}
\end{eqnarray}
\end{widetext}
where $R=r_{AB}$, $E_R = E_0 - 1/R$ and $Q_i$ are defined in Table \ref{TBL1}. 
These operators agree with those obtained previously for helium in the $R\rightarrow 0$ limit, as they should.

\paragraph{Gaussian integrals}
Almost all the calculations of matrix elements with $\alpha^6\,m$ operators are performed 
in this work by using the explicitly correlated Gaussian (ECG) functions. 
\begin{equation}
\phi_{\Sigma^+} = \Bigl(1+\frac{r_{12}}{2}\Bigr)
\,e^{-a_{1A}\,r_{1A}^2 -a_{1B}\,r_{1B}^2 -a_{2A}\,r_{2A}^2 -a_{2B}\,r_{2B}^2 - a_{12}\,r_{12}^2 } \label{24}
\end{equation}
In order to satisfy the electron-electron cusp condition, we include an explicit factor
$(1+r_{12}/2)$ in the wave function. It not only improves the numerical convergence, 
but also it is crucial for obtaining a correct numerical value for some of the nearly singular matrix elements,
especially $E'_A$, otherwise the result would be incorrect.
The other second-order matrix element $E_C$ does not have any singularities,
so the $(1+r_{12}/2)$ factor can be dropped. It involves intermediate states 
of $\Sigma^-$ and $\Pi$ symmetries, which have the following representations:
$\phi_{\Sigma^-} = \vec R \cdot (\vec r_{1A} \times \vec r_{2A}) \, \phi_{\Sigma^+}$ and
$\vec \phi_{\Pi} = (\vec R \times \vec r_{1A})\,\phi_{\Sigma^+}$.

The primary advantage of ECG functions is that all
integrals with operators in Table \ref{TBL1}, as well as in the second-order elements, 
can be performed either analytically or numerically as follows.
All the matrix elements are expressed as a linear combination of the following integrals
\begin{eqnarray}
f(n_1,n_2,n_3,n_4,n_5) &=& \frac{1}{\pi^3}\int d^3 r_1 \int d^3 r_2 \, 
r_{1A}^{n_1} r_{1B}^{n_2} r_{2A}^{n_3} r_{2B}^{n_4} r_{12}^{n_5}
\nonumber \\ && \hspace*{-10ex}
\times e^{-c_{1A}\,r_{1A}^2 -c_{1B}\,r_{1B}^2 -c_{2A}\,r_{2A}^2 -c_{2B}\,r_{2B}^2 - c_{12}\,r_{12}^2 } 
\label{27}
\end{eqnarray}
The ECG integrals with even powers of inter-particle distance can be generated by differentiation
over nonlinear parameters of the master integral
\begin{eqnarray}
f(0,0,0,0,0) = A^{-3/2} e^{-R^2 \frac{B}{A}} \label{29}
\label{basis_int}
\end{eqnarray}
where
\begin{eqnarray}
A &=& (c_{1A} + c_{1B} + c_{12}) (c_{2A} + c_{2B} + c_{12}) - c_{12}^2 \label{30}\\
B &=&   (c_{1B} + c_{1A}) c_{2A} c_{2B} + c_{1A} c_{1B} (c_{2A} + c_{2B}) 
\nonumber \\ &&
+ c_{12} (c_{1A} + c_{2A}) (c_{1B} + c_{2B}) \label{31}
\end{eqnarray}
If one of the $n_k$ indices is odd, the ECG integrals can also be obtained analytically by differentiation
of other master integrals. An exemplary master integral for the case of $n_1 = -1$ reads
\begin{equation}
f(-1,0,0,0,0) = \frac{1}{A \sqrt{A_1}}\,e^{-R^2 \frac{B}{A}}
F \bigg[R^2\bigg( \frac{B_1}{A_1} - \frac{B}{A}\bigg) \bigg], \label{32}
\end{equation}
where $A_1 = \partial_{c_{1A}} \, A$, $B_1 = \partial_{c_{1A}} \,B$, and $F(x) = {\rm erf}(x)/x$.
Molecular ECG integrals, as opposed to the atomic case, have no known analytic 
form when two or more $n_k$ are odd. In this case we use numerical integration 
with the quadrature adapted to the end-point logarithmic singularity \cite{gauslog}. 
This approach appears to be  very efficient for all the integrals with two and three odd indices,
which are required in the evaluation of matrix elements of all the $\alpha^6\,m$ operators. 

\paragraph{Numerical calculations}

The nonrelativistic wave function $\Phi$ used for the ground electronic state is the symmetrized
$(1\leftrightarrow 2, A\leftrightarrow B)$ linear combination of $N=128$, $256$, or $512$ basis functions 
$\phi_{\Sigma^+}$ from Eq. (\ref{24}). All individual nonlinear parameters are carefully optimized, and 
the precision achieved for the ground state energy is about $10^{-12}$ with $N=512$ basis.
The separate optimization with the same basis size $N$ was performed to accurately represent 
the resolvent of $\Pi$ and $\Sigma^{-}$ symmetry in the second-order matrix elements $E_C$.
Moreover, for $E_A^{'}$ we use an additional non-optimized constant sector of $\phi_{\Sigma^+}$ 
basis functions, where non-linear parameters come from the $\Sigma^+$  wave function of the size $N/2$. 
This is because the electronic ground state has to be subtracted from the resolvent.
The global optimization of all nonlinear parameters ensures high accuracy for matrix elements.
Nevertheless, in some cases, like for $Q_{10}$, we transform matrix elements
to a more regular but equivalent form to further improve the numerical accuracy \cite{drachman}. 
Moreover, for $Q_{27}$, $Q_{28}$, and $Q_{30}$ operators it was essential to use the basis functions
with $1+r/2$ prefactor, so the wave function satisfies exactly the electron-electron cusp
condition. Particular attention should be paid to the second-order matrix element $E'_A$
with the regularized Breit-Pauli Hamiltonian. 
The use of $1+r/2$ prefactor was necessary to subtract the $\delta^3(r)$ term from
the $\vec\nabla_1^2\,\vec\nabla_2^2$ differentiation of the outer wave function,
and it also significantly improves the numerical convergence of $E'_A$.

All numerical matrix elements have been checked against the $R\rightarrow 0$ and
$R\rightarrow\infty$ limits.  Every operator $Q_{i}$ in Table~\ref{TBL1} as well as 
$E_A^{'}$, $E_C$, and $E_{R2}$ have a well-defined limit $R \rightarrow 0$ to the corresponding 
helium ground state mean value \cite{hsinglet}. 
However, in the particular case of $\Pi$ contribution to $E_C$ the helium limit is achieved
at extremely low values of $R$, indicating the significant change in the character of
the electronic wave function at distances $R=0-0.2$ where the $E_C(\Pi)$ curve has a local sharp minimum.
The exceptional case is $E_{R1}$, which does not go to the helium limit at $R=0$.
The reason for this is that $Z\,\alpha$ expansion of the electron self-energy
assumes that all inter-particle distances are of the order of the Bohr radius.
When the  inter-nuclear distances are of the order of the electron Compton wavelength 
the $Z\,\alpha$ expansion takes a different form and the proper helium limit is then achieved. 

All the numerical matrix elements have also been verified against the long-distance 
asymptotics $R \rightarrow \infty$, which coincide with hydrogenic values as they should.
It was essential to perform all possible tests, in order to avoid mistakes
in derivation and coding of matrix elements. Moreover, matrix elements of  $Q_1 \ldots Q_7$
have also been calculated with the double James-Coolidge basis \cite{bo_h2} because the achieved 
numerical accuracy with exponential functions is much higher than with Gaussians. 
So far, we have not been able to calculate all the matrix elements with explicitly 
correlated exponential functions because they involve integrals that are too complicated,
but we plan to do this in the near future.   

\paragraph{Results}
The exemplary expectation value at $R=1.4$ au of all $Q_i$ operators is presented in Table \ref{TBL1}.
\begin{table}[htb]
%\begin{minipage}{18.0cm}
\renewcommand{\arraystretch}{0.95}
\caption{Expectation values of operators entering $H^{(6)}$ for
the $^1\Sigma^+$ state at $R=1.4$ au. The last digit in $Q_{9\ldots 28}$ is uncertain. The following notation was used in the table: 
$\vec r = \vec r_{12} = \vec r_1 - \vec r_2  $, 
$V_i = 1/r_{iA} + 1/r_{iB}$, $\vec V_i = \vec r_{iA}/r_{iA}^3 + \vec r_{iB}/r_{iB}^3$,
$\vec{P}=\vec{p}_1+\vec{p}_2$.  }
\label{TBL1}
%\begin{tabular}{l@{\hspace{0.2cm}}.@{\hspace{1.0cm}}.}
%\begin{tabular}{rl@{\hspace{0.1cm}}.}
\begin{ruledtabular}
\begin{tabular}{rl@{\hspace{0.1cm}}w{4.12}}
\rule[-1mm]{0mm}{4mm}&\multicolumn{1}{l}{Operator} &\multicolumn{1}{c}{Expectation value}  \\
\hline\\[-2ex]
$Q_{1}=$ &$ 4\,\pi\,\delta^{3}(r_{1A})         $& 2.888\,179\,88(1) \\
$Q_{2}=$ &$ 4\,\pi\,\delta^{3}(r)              $& 0.210\,402\,25(1) \\
$Q_{3}=$ &$ 4\,\pi\,\delta^{3}(r_{1A})/r_{2A}  $& 2.203\,14 \\
$Q_{4}=$ &$ 4\,\pi\,\delta^{3}(r_{1A})/r_{2B}  $& 2.778\,58 \\
$Q_{5}=$ &$ 4\,\pi\,\delta^{3}(r_{1A})\,p_2^2  $& 2.952\,30 \\
$Q_{6}=$ &$ 4\,\pi\,\delta^{3}(r)\,V_1         $& 0.604\,74 \\
$Q_{7}=$ &$ 4\,\pi\delta^{(3)}(r)\,P^2         $& 0.859\,90 \\
$Q_{8}=$ &$ 1/r                                $& 0.587\,36 \\
$Q_{9}=$ &$ 1/r^2                              $& 0.517\,93 \\
$Q_{10}=$&$ 1/r^3                              $& 0.414\,34 \\
$Q_{11}=$&$ V_1^2                              $& 4.852\,07 \\
$Q_{12}=$&$ V_1\,V_2                           $& 3.265\,50 \\
$Q_{13}=$&$ V_1/r                              $& 1.193\,32 \\
$Q_{14}=$&$ V_1\,V_2/r                         $& 2.454\,64 \\
$Q_{15}=$&$ V_1^2\,V_2                         $& 8.525\,27 \\
$Q_{16}=$&$ V_1^2/r                            $& 3.445\,33 \\
$Q_{17}=$&$ V_1/r^2                            $& 1.195\,29 \\
$Q_{18}=$&$\vec V_1\cdot \vec r/r^3            $& 0.406\,32 \\ 
$Q_{19}=$&$ \vec V_1\cdot \vec r/r^2           $& 0.488\,59 \\
$Q_{20}=$&$ V_1^i\,V_2^j\,(r^i r^j - 3\,\delta^{ij}\,r^2)/r 
                                               $& 0.547\,86 \\
$Q_{21}=$&$ p_2^2\,V_1^2                       $& 5.186\,77 \\
$Q_{22}=$&$ \vec p_1\, V_1^2\, \vec p_1        $& 5.145\,61 \\
$Q_{23}=$&$ \vec p_1\, /r^2\, \vec p_1         $& 0.554\,62 \\
$Q_{24}=$&$ p_1^i\,V_1\,(r^i\,r^j + \delta^{ij}\, r^2)/r^3\, p_2^j            
                                               $& 0.237\,37 \\
$Q_{25}=$&$ P^i\,(3\,r^i\,r^j - \delta^{ij} r^2)/r^5\,P^j                           
                                               $& -0.190\,40 \\
$Q_{26}=$&$  p_2^k\, V_1^i\,(\delta^{jk}\, r^i/r - \delta^{ik}\, r^j/r 
$\\&$
           - \delta^{ij}\, r^k/r - r^i\, r^j\, r^k/r^3)\,p_2^j
                                               $& -0.112\,60 \\
$Q_{27}=$&$ p_1^2\,p_2^2                       $& 1.328\,10 \\
$Q_{28}=$&$ p_1^2\,V_1\,p_2^2                  $& 5.208\,25 \\
$Q_{29}=$&$ \vec p_1\times\vec p_2\,/r\,\vec p_1\times\vec p_2           
                                               $& 0.386\,62 \\
$Q_{30}=$&$ p_1^k\,p_2^l\,(-\delta^{jl}\,r^i\,r^k/r^3 -
           \delta^{ik}\,r^j\,r^l/r^3 
$\\&$
            + 3\,r^i\,r^j\,r^k\,r^l/r^5)\, p_1^i\,p_2^j   
                                               $& -0.160\,82 \\
\end{tabular}
\end{ruledtabular}
%\end{minipage}
\end{table}
The numerical accuracy is about five significant digits, and we observe a significant cancellation,
so the sum, as expressed by $E_Q$, is smaller than most of the individual terms, see Table \ref{TBL2}.
The overall dependence of the non-logarithmic photon exchange contribution $E_Q+E_A+E_C+E_H = E'_Q + E'_A+E_C+E'_H$
on the inter-nuclear distance is presented in Fig. 1.
\begin{figure}%[!htb]
\caption{Non-logarithmic photon exchange contribution $E'_Q + E'_A+E_C+E'_H$ as a function of the inter-nuclear distance $R$.
         The horizontal line is located at $-1/8$, which is twice the atomic hydrogen value, and
         the dashed curve shows the $0.529\,947\,904/R^2-1/8$ asymptotics, which is obtained 
         from the small $R$ expansion of the Casimir-Polder potential \cite{lach}.}
\includegraphics[scale=0.5]{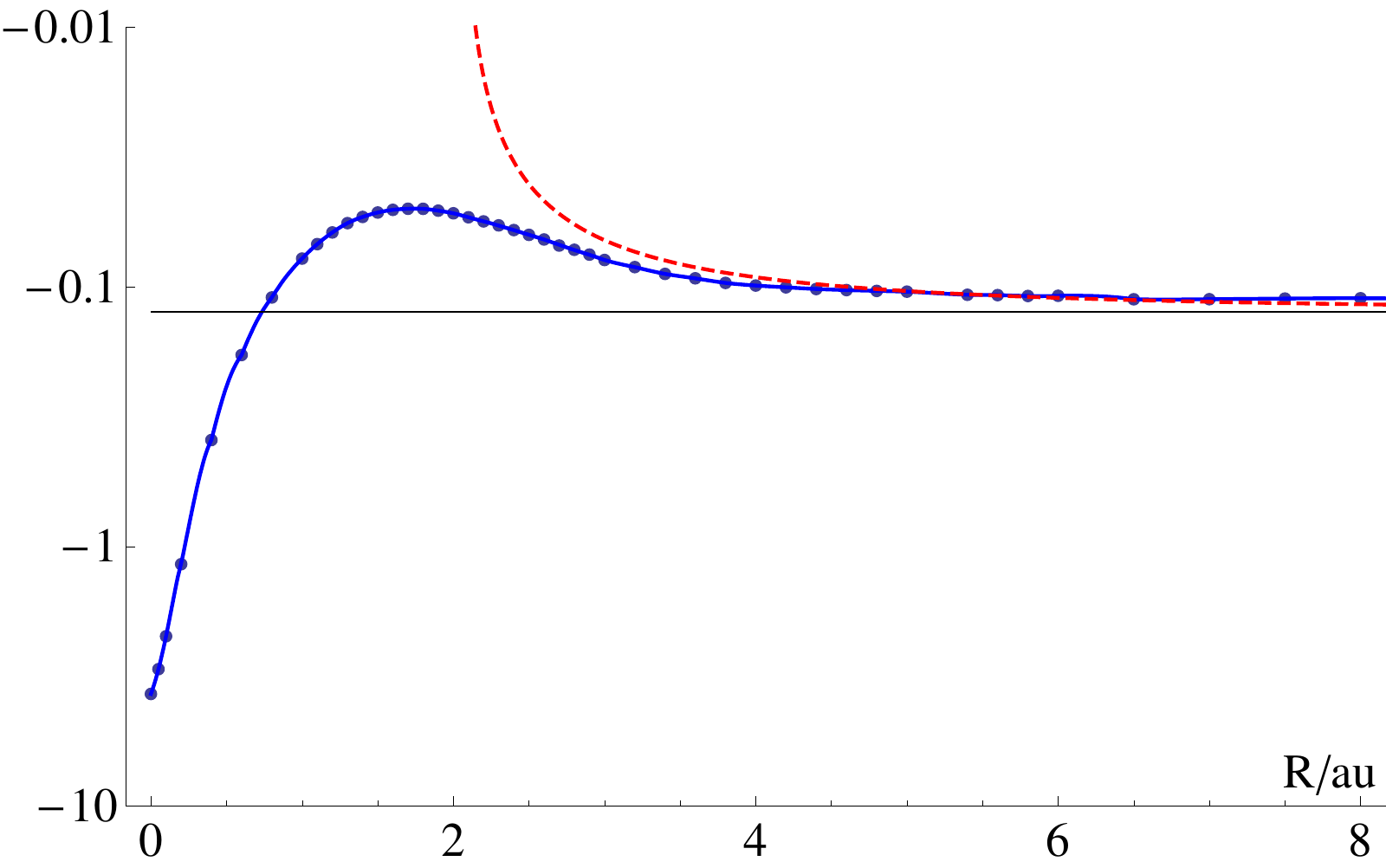}
\vspace*{-5ex}
\end{figure}
We observe the minimum around 1.5 au, which is not far from the mean internuclear distance 
where the radial wave function is localized, so the photon exchange contribution to the dissociation energy
is relatively small.

Table \ref{TBL2} supplies all contributions to $E^{(6)}$ as given in Eq. (\ref{21}) at $R=1.4$ au.
It is worth noting that the by far largest contribution comes from
the one-loop radiative correction $E_{R1}$, which legitimizes the previous estimations
for $\alpha^6\,m$ contribution \cite{piszcz}.
Table \ref{TBL3} presents a summary of all contributions to the dissociation,
fundamental vibrational and rotational transitions. In particular, this table contains 
significantly more accurate results for the $\alpha^2\,m$ nonrelativistic energies obtained
using explicitly correlated exponential functions~\cite{nonad}. 
\vspace*{-2ex}
\begin{table}[h]
\renewcommand{\arraystretch}{1.0}
\caption{Contributions to $E^{(6)}$ for the ground electronic state of H$_2$ at $R=1.4$ au.
$E_{LG}$ is the logarithmic correction, last term in Eq. (\ref{21}), $E_D$ is the $\alpha^6\,m$ contribution
from the Dirac equation.}
\label{TBL2}
\begin{ruledtabular}
\begin{tabular}{c@{\extracolsep{\fill}}w{3.11}w{3.11}}
$\alpha^6\,m$         & \centt{H$_2(\Sigma^+$)} \\
\hline\\[-2ex]
$E'_Q$                &  0.688\,40(16)   \\
$E'_H$                & -0.043\,832   \\
$E'_A$                & -0.641\,4(5)   \\
$E_C$                 & -0.059\,54(4)   \\
\hline
Subtotal              & -0.056\,4(6)   \\ 
$E_{R1}$              &  9.254\,583    \\
$E_{R2}$              &  0.142\,233    \\
$E_{LG}$              &  0.258\,811      \\
\hline
Total                 &  9.599\,3(6)    \\[1ex]
$-2\,E_D({\rm H})$    &   0.125\,000      \\
$-2\,E_{R1}({\rm H})$ &  -6.123\,245        \\
$-2\,E_{R2}({\rm H})$ &  -0.109\,212    \\
\hline
$E^{(6)}({\rm H}_2) - 2\,E^{(6)}({\rm H})$   &  3.491\,8(6)\,\alpha^6\,m
\end{tabular}
\end{ruledtabular}
\end{table}
                                         
\paragraph{Summary}
We have calculated the complete $\alpha^6\,m$ contribution to the molecular hydrogen energy levels.
This is the first calculation of the higher order relativistic effects ever performed for molecules,
except for the one-electron molecular ion H$_2^+$ \cite{korobov}.
Besides significant improvements in the H$_2$ levels, it shows how to properly incorporate electron 
correlations with relativistic and QED effects. 
\begin{table}[!htb]
\renewcommand{\arraystretch}{1.3}
\caption{Contributions to dissociation, vibrational, and rotational transitions in H$_2$ in cm$^{-1}$. 
         Physical constants are from \cite{nist} and $r_p = 0.8409(4)$ fm.
         There are additional $10^{-3}$ uncertainties of $\alpha^4$, $\alpha^5$, 
         and $\alpha^6\,m$ terms due to the BO approximation,which are included
         in the final result only. 
         $\alpha^7\,m$ term is estimated from the known leading double logarithmic
         contribution in Eq.~(\ref{14}) and the related 50\% uncertainty is assumed.
         $E_{r_p^2}$ is the finite proton size correction.}
\label{TBL3}
\begin{ruledtabular}
\begin{tabular}{l@{\extracolsep{\fill}}w{6.12}w{3.11}w{2.10}}
             & \cent{D_0} & \cent{v=0\rightarrow 1} & \cent{J=0\rightarrow 1} \\
\hline
$\alpha^2\,m$ & 36\,118.797\,746\,12(5)& 4\,161.164\,070\,3(1) & 118.485\,260\,46(3) \\
$\alpha^4\,m$ &      -0.531\,8(3)^{a}    &      0.023\,41(1)^{c}   &   0.002\,580(1)  \\
$\alpha^5\,m$ &      -0.194\,8(2)^{b}    &     -0.021\,29(2)^{c}   &  -0.001\,022(1)  \\
$\alpha^6\,m$ &      -0.002\,065(6)    &     -0.000\,192\,3(6)     &  -0.000\,008\,9(1) \\
$\alpha^7\,m$ &       0.000\,118(59) &      0.000\,012\,0(60) &   0.000\,000\,6(3) \\
$E_{r_p^2}$   &      -0.000\,031       &     -0.000\,003\,2    &  -0.000\,000\,2     \\ 
\hline
Theory                & 36\,118.069\,1(6)   & 4\,161.166\,01(4)  & 118.486\,810(4)   \\
\cite{exp1,exp2,exp3} & 36\,118.069\,62(37) & 4\,161.166\,32(18) & 118.486\,84(10)   \\
\end{tabular}                            
\end{ruledtabular}     
$^a$ \cite{kom11}; $^b$ \cite{piszcz};$^c$ \cite{exp2}.\hfill\
\end{table}                              

The improvement of the H$_2$ levels down to the $10^{-7}$ cm$^{-1}$ level will lead to more accurate 
determination of the R$_\infty$ constant and may shed light on the proton charge radius puzzle.
The ratio of the nuclear finite size effects to the transition energy for $1S-2S$ in H is $3.9\cdot 10^{-10}$,
while for the H$_2$ dissociation energy it is $8.6\cdot 10^{-10}$. Since the ratios are sufficiently different,
one can use these transitions to determine R$_\infty$ and $r_p$ without referring to 
the other, less well-known transitions in hydrogen. To achieve this, however, further improvement
in H$_2$ levels is required, in particular the calculation of the $\alpha^7\,m$ contribution. 

\begin{acknowledgments}
We wish to thank Grzegorz \L ach for his interesting discussions, and for the calculation of $1/R^2$
asymptotics and the fit of $E^{(6)}(R)$. This work was supported by the National Science Center (Poland) Grant
Nos. 2012/04/A/ST2/00105 (P.C. and K.P.) and 2014/13/B/ST4/04598 (M.P. and J.K.), 
as well as by a computing grant from the Poznan Supercomputing and Networking Center, 
and by PL-Grid Infrastructure.
\end{acknowledgments}

\end{document}